\documentclass[11pt]{article}
\usepackage{moriond,epsfig}
\usepackage{amssymb}
\usepackage{amsbsy}

\def\be{\begin{equation}}
\def\ee{\end{equation}}
\def\bea{\begin{eqnarray}}
\def\eea{\end{eqnarray}}

\renewcommand{\r}{{\boldsymbol r}}

\newcommand{\q}{{\boldsymbol q}}

\newcommand{\ep}{\epsilon}

\begin{document}
%%%%%%%% HAS BEEN SUPPRESSED \vspace*{4cm}
\title{Non-exponential relaxations in disordered conductors}

\author{\underline{Gilles Montambaux}$^1$,
Eric Akkermans$^2$}

\address{$^1$Laboratoire de Physique des Solides, CNRS UMR 8502, \\
  Universit\'e Paris-Sud, 91405 Orsay, France}
\address{$^2$Department of Physics, Technion Israel Institute
of Technology, 32000 Haifa, Israel}

\maketitle\abstracts{ We show that, in low dimensional conductors,
the  quasiparticle decay and the relaxation of the phase are not
exponential processes. In quasi-one dimension, they scale as $e^{-
(t/\tau_N)^{3/2}}$ where the characteristic time $\tau_{in}$,
identical for both processes, is a power $T^{2/3}$ of the
temperature. This result implies a distribution of
relaxation times.}

The problem of dephasing in the presence of electron-electron
interaction in disordered conductors was first addressed in the
pioneering work of Altshuler and Aronov \cite{Altshuler8513}. It
has been revived by a  series of recent experiments
\cite{Mohanty9713,Gougam99,Esteve02}. Another very related
question is the understanding of the time evolution of a quasiparticle
state which governs the relaxation towards equilibrium. This
quasiparticle decay is usually described by a characteristic time,
implying implicitly an exponential process. Here we show that in
low dimension, and in particular in quasi-one-dimensional
disordered wires, this decay is   faster than exponential. We
focus on two aspects, the decay of a quasiparticle state and
the loss of phase coherence. It has been shown that both processes
are related \cite{Imry}. They are described by
the same characteristic time scale. Here we show that {\it both}
processes are non exponential \cite{AkkMon04,Livre}.
\medskip

 We start by considering
the lifetime of a quasiparticle state, initially  in a state of energy
$\ep$. Its interaction with another quasiparticle of energy $\ep'$
leads to final states at energies $\ep -\omega$ and
$\ep'+\omega$.   At zero temperature, the lifetime is given by
%%%%%%%%%%%%%%%%%%%%%%%%%%%%%%%%%%%%%%%%%%%%%%%%%%%%%%%%%%%
\be
  {1 \over \tau_{ee}(\ep )}= {4 \pi \nu_0^3}
\int_0^\ep \omega K(\omega) d \omega \label{taugenW2} \ee where
$\nu_0$ is the density of states per spin direction. The kernel
$K(\omega)$ is the disordered averaged squared matrix element of
the screened Coulomb interaction. It depends only on the energy
transfer $\omega$ and it can be written as \cite{Altshuler8513}

%%%%%%%%%%%%%%%%%%%%%%%%%%%%%%%%%%%%%%%%%%%%%%%%%%%%%%%%%%%
 \be K(\omega)={1 \over \pi^2 \nu_0^2 \Omega^2
}\sum_{\q \neq 0} |U(\q,\omega)|^2 [ \mbox{Re}P_d(\q,\omega)]^2 \
\ . \label{W2q20}\ee
%%%%%%%%%%%%%%%%%%%%%%%%%%%%%%%%%%%%%%%%%%%%%%%%%%%%%%%%%%%
In this expression $P(\q,\omega)=1/(- i \omega + D q^2)$ is the
classical diffusion probability,  $U(\q,\omega)$ is the
dynamically screened coulomb potential $U(\q,\omega)= (1 /2
\rho_0) D q^2/(- i \omega + D q^2)$ and $\rho_0$ is the density of
states per spin direction and per volume unit. The kernel can also
be written under the form
%%%%%%%%%%%%%%%%%%%%%%%%%%%%%%%%%%%%%%%%%%%%%%%%%%%%%%%%%%%
\be K(\omega)=    {1 \over 4 \pi^2 \nu_0^4}  \sum_\q {1 \over
\omega} \mbox{Im}P_d(\q,\omega) ={1 \over 4 \pi^2 \nu_0^4
}\sum_{\q \neq 0} {1 \over \omega^2 + D^2 q^4} \ \ ,
\label{Weeq21} \ee
%%%%%%%%%%%%%%%%%%%%%%%%%%%%%%%%%%%%%%%%%%%%%%%%%%%%%%%%%%%
so that the lifetime reduces to
%%%%%%%%%%%%%%%%%%%%%%%%%%%%%%%%%%%%%%%%%%%%%%%%%%%%%%%%%%%
 \be {1 \over
\tau_{ee}(\ep)} ={1 \over\pi \nu_0 } \int_0^\ep   d \omega \sum_\q
\mbox{Im}P_d(\q,\omega) \ \ . \label{taueeq21}\ee
%%%%%%%%%%%%%%%%%%%%%%%%%%%%%%%%%%%%%%%%%%%%%%%%%%%%%%%%%%%
Replacing the discrete sum over $\q$ by an integral leads to

\begin{equation} K(\omega)= {\alpha_d \over 16 \nu_0^4
\omega^2 }\left( {\omega \over
E_c} \right)^{d/2}\\
\label{W2q33}
\end{equation}
where $\alpha_1= \sqrt{2}/\pi^2$, $\alpha_2= 1/2\pi^2$,
$\alpha_3=\sqrt{2}/2\pi^3 $ and $E_c=D/L^2$ is the Thouless
energy. From Eq. (\ref{taugenW2}), one obtains the well-known
\cite{Altshuler8513} power-law dependence of the lifetime
$1/\tau_{ee}(\ep) \propto \ep^{d/2}$.
\medskip

At first sight, one could consider that,   at finite
temperature, the quasiparticle relaxation rate is given by $1 /
\tau_{ee}(\ep,T) \propto \mbox{max}(\ep^{d/2},T^{d/2})$. This is
not correct for $d \leq 2$ because of the importance of the processes
with low energy transfer \cite{Altshuler8513}. However, it has
been found that the quasiparticle relaxation is still exponential
with a characteristic time  which varies like $T^{2/3}$ in one
dimension \cite{Altshuler8513}. Here we show that in low
dimensions $d \leq 2$, {\it the relaxation is not
exponential}. To that purpose, we start from the expression

%%%%%%%%%%%%%%%%%%%%%%%%%%%%%%%%%%%%%%%%%%%%%%%%%%%%%%%%%%%
\be {1 \over \tau_{ee}(\ep,T)}= 4 \pi \nu_0^3
\int_{-\infty}^\infty d \omega \int_{-\infty}^\infty d \ep'
F(\ep,\ep',\omega) K(\omega) \label{tau6} \ee
%%%%%%%%%%%%%%%%%%%%%%%%%%%%%%%%%%%%%%%%%%%%%%%%%%%%%%%%%%%
that generalizes Eq. (\ref{taugenW2}) to finite temperatures. The
thermal function $F(\ep,\ep',\omega)$ depends on the Fermi factors
$f_\ep=1/(e^{\beta \ep}+1)$ and has the form
%%%%%%%%%%%%%%%%%%%%%%%%%%%%%%%%%%%%%%%%%%%%%%%%%%%%%%%%%%%
\be F(\ep,\ep',\omega)= f_{\ep'}
(1-f_{\ep-\omega})(1-f_{\ep'+\omega})
+(1-f_{\ep'})f_{\ep-\omega}f_{\ep'+\omega}\ \ . \label{Ft} \ee
%%%%%%%%%%%%%%%%%%%%%%%%%%%%%%%%%%%%%%%%%%%%%%%%%%%%%%%%%%%
 The first term is larger  for $\ep >0$, it describes the relaxation of an
electronic state above the Fermi level. The second term, larger
for $\ep<0$, accounts for the relaxation of a  hole state within
the Fermi sea. The two terms are equal for $\ep=0$. Integrating
over $\ep'$, and after some straightforward manipulations, it rewrites \be{1 \over \tau_{ee}(\ep,T)}= 2 \pi \nu_0^3
\int_{-\infty}^\infty d \omega K(\omega)\omega\left( \coth {\beta
\omega \over 2} +\tanh{\beta\over 2} (\ep-\omega)\right)\ \ .
\label{tauab}\ee
%%%%%%%%%%%%%%%%%%%%%%%%%%%%%%%%%%%%%%%%%%%%%%%%%%%%%%%%%%%
This expression can also be obtained from the imaginary part of
the self-energy,  a calculation that  incorporates implicitly both
processes involved in  Eq. (\ref{Ft}) \cite{Abrahams81}.
$\tau_{ee}(\ep,T)$ is defined as the lifetime of a quasiparticle.
It can   also be interpreted as being the characteristic time that
describes the relaxation towards equilibrium. Indeed, it can be
defined from a Boltzmann equation for the energy distribution
$n_\ep$ \cite{Altshuler8513,Schmid74}
%%%%%%%%%%%%%%%%%%%%%%%%%%%%%%%%%%%%%%%%%%%%%%%%%%%%%%%%%%%
\be \label{boltzfe1}
 {\partial n_\ep \over
\partial t}= - 4 \pi \nu_0^3 \int_{-\infty}^\infty
d \omega  K(\omega)    \int_{-\infty}^\infty d \ep'
[n_{\ep}n_{\ep'} (1-n_{\ep-\omega})(1-n_{\ep'+\omega})
  -n_{\ep-\omega}n_{\ep'+\omega}
(1-n_{\ep})(1-n_{\ep'})] \ee
%%%%%%%%%%%%%%%%%%%%%%%%%%%%%%%%%%%%%%%%%%%%%%%%%%%%%%%%%%%
The relaxation term contains two contributions that respectively
describe the quasiparticles leaving a given quantum state ( "out"
term) and incoming in this state  ("in"). At equilibrium $n_\ep$
equals the Fermi factor $f_\ep= 1/(e^{\beta \ep}+1)$ and the
relaxation term is zero. Linearising around the equilibrium
distribution $n_\ep= f_\ep + \delta n_\ep$, one obtains a relaxation
of the form    $ {\partial \delta n_\ep  /
\partial t}=- {\delta n_\ep / \tau_{ee}(\ep,T)}$ with the
same characteristic time given by (\ref{tau6}). The lifetime of a quasiparticle
state
can  also be interpreted as the relaxation time of the energy
distribution.

\medskip

We now consider the lifetime of a quasiparticle {\it at the Fermi
level} and we denote it by $\tau_{in}(T)$. From Eq. (\ref{tauab}), it
is given by
 \cite{Eiler}
%%%%%%%%%%%%%%%%%%%%%%%%%%%%%%%%%%%%%%%%%%%%%%%%%%%%%%%%%%%
\be {1 \over \tau_{in}(T)}={1 \over \tau_{ee}(0,T)}= {8 \pi
\nu_0^3} \int_0^\infty d \omega K(\omega) {\omega \over \sinh
\beta \omega} \label{tauwbeta} \ee
%%%%%%%%%%%%%%%%%%%%%%%%%%%%%%%%%%%%%%%%%%%%%%%%%%%%%%%%%%%
In $d=3$, the integral converges, leading to the well-known
$T^{3/2}$ behavior \cite{Altshuler8513}. For $d \leq 2$, the
integral diverges at low energy transfer. It has been argued that
the integral has to be cut-off at low energy: since the lifetime
of a quasiparticle is finite, no energy transfer can be  smaller
than this inverse lifetime \cite{Altshuler8513}. Consequently, the
lifetime is solution of the self-consistent equation (for
simplicity the thermal factor in Eq. (\ref{tauwbeta}) has been
replaced by an upper cut-off at  $\omega \sim T$) :

%%%%%%%%%%%%%%%%%%%%%%%%%%%%%%%%%%%%%%%%%%%%%%%%%%%%%%%%%%%
\be {1 \over  \tau_{in}(T)} \simeq {  T   \over \nu_0} \int_{1/
\tau_{in}(T)}^T {d \omega  \over \omega^2} \left( {\omega \over
E_c}\right)^{d/2} \ \ . \label{tauphi1D1}\ee
%%%%%%%%%%%%%%%%%%%%%%%%%%%%%%%%%%%%%%%%%%%%%%%%%%%%%%%%%%%
 This argument
gives a  characteristic time $\tau_{in}(T)$ which scales as
$T^{2/3}$ in one dimension :

 %%%%%%%%%%%%%%%%%%%%%%%%%%%%%%%%%%%%%%%%%%%%%%%%%%%%%%%%%%%
\be {1 \over  \tau_{in}(T)}\sim \left({\Delta T \over
E_c^{1/2}}\right)^{2/3}\ \ . \label{tauphi1D4}\ee
%%%%%%%%%%%%%%%%%%%%%%%%%%%%%%%%%%%%%%%%%%%%%%%%%%%%%%%%%%%
$\Delta=1/\nu_0$ is the interlevel energy spacing.

\medskip

We argue that the   divergence at low energy transfer in
(\ref{tauphi1D1}) has a deeper significance \cite{AkkMon04}. It is
the signature of a {\it non-exponential relaxation}. The reason
goes as follows: the quasiparticle state decay cannot be
exponential since, after a time $t$, the energy transfer cannot be
defined with a precision better than   $1/t$ (Heisenberg
inequality). Thus {\it the decay rate cannot be constant in time}
\cite{AkkMon04}. Instead of assuming a relaxation of the inital
state of the form ${\cal P}(t)=e^{-t /\tau_{in}}$ with eq.
(\ref{tauphi1D1}), we find that the probability ${\cal P}(t)$ for
the quasiparticle to stay in its initial state is given by

%%%%%%%%%%%%%%%%%%%%%%%%%%%%%%%%%%%%%%%%%%%%%%%%%%%%%%%%%%%
\be -  \ln  { \cal P} \simeq { T t  \over \pi \nu_0} \int_{1/t}^T
{d \omega \over \omega^2} \left( {\omega \over E_c}\right)^{d/2} \
\ ,\label{tauphi1D2}\ee
%%%%%%%%%%%%%%%%%%%%%%%%%%%%%%%%%%%%%%%%%%%%%%%%%%%%%%%%%%%,
 an expression valid for times $t \gg 1/T$. In one  dimension, one has \cite{Livre}:
%%%%%%%%%%%%%%%%%%%%%%%%%%%%%%%%%%%%%%%%%%%%%%%%%%%%%%%%%%%
\be   -\ln{\cal P}  =  { T t   \over  \sqrt{2} \pi \nu_0}
\int_{1/t }^T {d \omega  \over \omega^2} \left( {\omega \over
E_c}\right)^{1/2}= {  \sqrt{2}  T \over \pi \nu_0 \sqrt{E_c}}
t^{3/2} \label{tauphi1D3}\ee
%%%%%%%%%%%%%%%%%%%%%%%%%%%%%%%%%%%%%%%%%%%%%%%%%%%%%%%%%%%
which leads to a non-exponential behavior for the relaxation of a quasiparticle
state.

\begin{equation}
\begin{array}{c}
\displaystyle {\cal P}(t,T) \sim e^{- [t/ \tau_{in}(T)]^{3/2}}\ \
\ \ \ \ \ d=1 \label{tauphi1D33}
\end{array}
\ee with the same characteristic time as in Eq. (\ref{tauphi1D4}).
%%%%%%%%%%%%%%%%%%%%%%%%%%%%%%%%%%%%%%%%%%%%%%%%%%%%%%%%%%%,
Similarly, in two dimensions, starting from (\ref{tauphi1D2}),
one finds a logarithmic correction to the exponential decrease \be
{\cal P}(t,T) \sim e^{-{t \over \tau_{in}}{ 1 \over  \ln T t}}
\qquad d=2 \label{tauphi2D33}\ee where $\tau_{in}(T) \propto
\Delta T/E_c$.

\medskip

Now, one can wonder whether this peculiar behavior of the relaxation
of a quasiparticle state appears in the time
dependence of other quantities. To that purpose, we consider the phase coherence
time. It is defined as the lifetime of the Cooperon in the
presence of other electrons. The Cooperon is the quantum
correction to the return probability that consists in pairs of time
reversed trajectories. Altshuler, Aronov and Khmelnitskii \cite{Altshuler8213} have
shown that the effect of other electrons can be accounted for by a
fluctuating electric potential $V(\r,\tau)$ whose characteristics
are given by the fluctuation-dissipation theorem:

 \be \langle V(\r,\tau)V(\r',\tau')\rangle_T={\delta(\tau-\tau')  \over (2
 \pi)^d}{2 T \over \sigma_0} \int {d\q \over q^2}e^{i \q. (\r-\r')}
 \label{VVrr'} \ee
The Cooperon contribution to the
return probability takes the form

\be  P_c(\r,\r,t)=   P_c^{(0)}(\r,\r,t) \left\langle e^{i
\Phi(\r,t) }\right\rangle_{T,{\cal C}} \label{Pcfluc1}  \ \ \  \ee
where $P_c^{(0)}$ is the cooperon  in the absence of the
fluctuating  potential.
  $\Phi=\Phi(\r,t)$ is the relative phase for a pair of time reversed trajectories
  at time $t$   :

%%%%%%%%%%%%%%%%%%%%%%%%%%%%%%%%%%%%%%%%%%%%%%%%%%%%%%%%%%%
\be \Phi ={e \over \hbar} \int_{0}^{t} [V(\r(\tau),\tau)-V(\r(
\tau),\overline{\tau})] d\tau \label{PhiVV}\ee
%%%%%%%%%%%%%%%%%%%%%%%%%%%%%%%%%%%%%%%%%%%%%%%%%%%%%%%%%%%
One of the trajectories propagates from time $\tau=0$ to $\tau=t$,
while the time reversed trajectory propagates from $\tau=t$ to
$\tau=0$. We define  $\overline{\tau}=t-\tau$ and
 $\langle \cdots \rangle_{T,{\cal C}}$ is the average taken both on the distribution of diffusion trajectories    ($\langle \cdots \rangle_{{\cal C}}$)
 and on the thermal fluctuations ($\langle \cdots \rangle_{T}$) of the
 electric potential. The thermal  fluctuations are gaussian so that the thermal average
   $\left\langle e^{i \Phi }\right\rangle _T $ satisfies
    :
%%%%%%%%%%%%%%%%%%%%%%%%%%%%%%%%%%%%%%%%%%%%%%%%%%%%%%%%%%%
 \be
\left\langle e^{i \Phi }\right\rangle_T = e^{- {1 \over 2}
\left\langle\Phi^2 \right\rangle_T}  \ \ .\ee
%%%%%%%%%%%%%%%%%%%%%%%%%%%%%%%%%%%%%%%%%%%%%%%%%%%%%%%%%%%
 Let us start with the calculation of $\left\langle \Phi^2  \right\rangle_T$. Inserting (\ref{VVrr'}) into (\ref{PhiVV}), one finds :

%%%%%%%%%%%%%%%%%%%%%%%%%%%%%%%%%%%%%%%%%%%%%%%%%%%%%%%%%%%

\begin{equation}
   \left\langle \Phi^2 \right\rangle_T={4 e^2 T \over
 \sigma_0 \hbar^2} \int_{0}^{t}d\tau
 \int {d\q \over (2 \pi)^d} { 1 \over q^2}[1 - \cos \q.
 (\r(\tau)-\r(\overline{\tau})) ] \ \ .
\label{PhiVV3}\ee
%%%%%%%%%%%%%%%%%%%%%%%%%%%%%%%%%%%%%%%%%%%%%%%%%%%%%%%%%%%
In one dimension, the phase fluctuation becomes
%%%%%%%%%%%%%%%%%%%%%%%%%%%%%%%%%%%%%%%%%%%%%%%%%%%%%%%%%%%
 \be \left\langle \Phi ^2 \right\rangle_T={2 e^2
T \over \hbar^2 \sigma_0 S} \int_{0}^{t}  | r(\tau) -
r(\overline{\tau})| d\tau \ \ .\label{Phi2Vrt} \ee
%%%%%%%%%%%%%%%%%%%%%%%%%%%%%%%%%%%%%%%%%%%%%%%%%%%%%%%%%%%
It depends on the trajectory  $r(\tau)$ and we still need to calculate the
 average
 %%%%%%%%%%%%%%%%%%%%%%%%%%%%%%%%%%%%%%%%%%%%%%%%%%%%%%%%%%
\be \left\langle e^{-{1 \over 2}\left\langle \Phi^2
\right\rangle_T} \right\rangle_{\cal C}\label{PhiVV4}\ee
%%%%%%%%%%%%%%%%%%%%%%%%%%%%%%%%%%%%%%%%%%%%%%%%%%%%%%%%%%%
over the distribution of diffusive trajectories.  A first
approximation consists in assuming  that  $\left\langle e^{-{1 \over
2}\left\langle \Phi^2 \right\rangle_T} \right\rangle_{\cal C}=
e^{-{1 \over 2}\left\langle \Phi^2 \right\rangle_{T,{\cal C}}}$.
Averaging over the diffusive trajectories, we obtain that $|r(\tau)-r(\overline{
\tau})|$ scales as $\sqrt t$ so that, from
(\ref{Phi2Vrt}), one has

 \be \left\langle \Phi^2
\right\rangle_{T,{\cal C}}=  {\sqrt{\pi}  e^2 T \over 2 \hbar^2
\sigma_0 S }\sqrt{ D}\  t^{3/2}  = {\sqrt{\pi} \over 2} \left({t
\over \tau_N}\right)^{3/2}\label{Phi2}\ee where we have defined
the characteristic time

%%%%%%%%%%%%%%%%%%%%%%%%%%%%%%%%%%%%%%%%%%%%%%%%%%%%%%%%%%%
\be \displaystyle \tau_N=\left({\hbar^2 \sigma_0 S \over   e^2 T
\sqrt{ D}}\right)^{2/3} \label{taunyquist} \ee
%%%%%%%%%%%%%%%%%%%%%%%%%%%%%%%%%%%%%%%%%%%%%%%%%%%%%%%%%%%
so that we find, within this approximation :

%%%%%%%%%%%%%%%%%%%%%%%%%%%%%%%%%%%%%%%%%%%%%%%%%%%%%%%%%%%
 \be
\left\langle e^{i \Phi }\right\rangle_{T,{\cal C}} \simeq
e^{-{\sqrt{\pi}/4}
 \left({t / \tau_N}\right)^{3/2}} \ \ . \label{gauss}\ee
A similar behavior has been found in \cite{Stern13}. Apart from a
numerical factor, the characteristic time (\ref{taunyquist})  is
the same as the inelastic time (\ref{tauphi1D4}) for the
decay of a quasiparticle state. Moreover Eq. (\ref{Phi2}) shows that the
phase relaxation has the same temporal dependence as the
quasiparticle relaxation (\ref{tauphi1D33}).

However, the result (\ref{gauss}) is not yet fully correct since
we have made an approximation in replacing the average $\langle
\cdots \rangle$ of the exponential by the exponential of the
average. The exact behavior of the phase relaxation (\ref{PhiVV4})
can be obtained from a functional integral approach
\cite{Altshuler8213} that shows that   the Laplace transform $$P_\gamma(r,r)=\int dt P_c(r,r,t)
e^{-\gamma t}$$
  is given by :
%%%%%%%%%%%%%%%%%%%%%%%%%%%%%%%%%%%%%%%%%%%%%%%%%%%%%%%%%%%
\be P_\gamma(r,r) =-{1 \over 2} \sqrt{\tau_N \over D}
{\mbox{Ai}(\tau_N/\tau_\gamma) \over
\mbox{Ai}'(\tau_N/\tau_\gamma)}\label{PgC2g} \ee
%%%%%%%%%%%%%%%%%%%%%%%%%%%%%%%%%%%%%%%%%%%%%%%%%%%%%%%%%%%
with  $\mbox{Ai}$ et $\mbox{Ai}'$ being respectively the Airy
function and its derivative  \cite{Abramovitz13} and
$\tau_\gamma=1/\gamma$. The probability $P_c(r,r,t)$ could be
obtained from inverse Laplace transform of Eq. (\ref{PgC2g}). Its
time dependence  is clearly non exponential. Here we  obtain the
time dependence  of the phase relaxation. It is of the form
$\left\langle e^{i \Phi} \right\rangle_{T,{\cal C}}=f(t /\tau_N)$.
Since  in one dimension, one has $P_c^{(0)}(\r,\r,t)=1/\sqrt{4 \pi
D t}$, the unknown function $f(t/\tau_N)$ is solution of

%%%%%%%%%%%%%%%%%%%%%%%%%%%%%%%%%%%%%%%%%%%%%%%%%%%%%%%%%%%
\be   \int_0^\infty {1 \over \sqrt{4 \pi D t}} f\left({t \over
\tau_N}\right)e^{-t/\tau_\gamma}dt=-{1\over 2}\sqrt{\tau_N \over
 D} {\mbox{Ai}(\tau_N/\tau_\gamma) \over
\mbox{Ai}'(\tau_N/\tau_\gamma)}\label{airy4}\ee
%%%%%%%%%%%%%%%%%%%%%%%%%%%%%%%%%%%%%%%%%%%%%%%%%%%%%%%%%%%
The inverse Laplace transform is obtained by noticing that both
the Airy function and its derivative are analytic and non
meromorphic functions whose zeroes lie on the negative real axis.
Then, by performing the integral in the complex plane with the
residues $\mbox{Res}(e^{st} \mbox{Ai}(s) / \mbox{Ai}'(s)) =
e^{-|u_{n}|t} / |u_{n}|$ where the $u_n$ are the zeros of
$\mbox{Ai}'(s)$ given at a very good approximation by $|u_{n}| =
\left( {3 \pi \over 2} (n - {3 \over4} )\right)^{2/3} $
\cite{Abramovitz13}, we obtain the analytic function \be
\left\langle e^{i \Phi} \right\rangle_{T,{\cal C}}= \sqrt{{\pi t
\over \tau_{in}}} \sum_{n=1}^\infty {e^{- |u_n| t/\tau_{in}} \over
|u_n|} \label{fun} \ee
 At small times $t < \tau_{in}$,  it   behaves like eq.
(\ref{gauss}). At large time, the relaxation is driven by the
first zero of the $\mbox{Ai}'$ function, namely $ \left\langle
e^{i \Phi} \right\rangle_{T,{\cal C}} \simeq \sqrt{\pi t /
\tau_{in}}
 e^{-|u_1| t/\tau_{in}} / |u_1| $ with $|u_1| \simeq 1.019$.
Clearly, the relaxation (\ref{fun}) is never exponential. It
appears as a  distribution of relaxation times $\tau_{in} /
|u_{n}|$ which is at the origin of the rather unexpected
compressed exponential behavior of the quasiparticle decay and of
the Cooperon phase relaxation.

\begin{figure}
\centerline{ \epsfxsize 8cm \epsffile{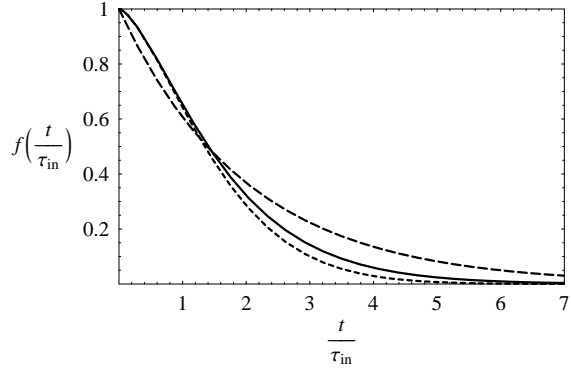} }
\caption{\it Behaviour of $ \langle e^{i \Phi(t)}\rangle_{T,{\cal
C}}$. The continuous line is the exact result (\ref{fun}). The
dotted line is obtained from the small time expansion
(\ref{gauss}). The dashed line shows the exponential fit $e^{-  t
/ 2\tau_{in}}$.} \label{airyplot}
\end{figure}

 In conclusion, we have shown that the decay
of a quasiparticle state in a low dimensional disordered conductor is
not exponential. In quasi-1d, it is of the form $e^{-(t
/\tau_{in})^{3/2}}$. We have also calculated the relaxation of the
phase in a quasi-1d conductor and we have shown that is it also
described by the same time dependence as for the relaxation of the
energy, with the same characteristic time.

This research was supported in part by the Israel Academy of
Sciences,  by the Fund for Promotion of Research at the
Technion, and by the French-Israeli Arc-en-ciel program.

\end{document}